\documentclass[pra,reprint,twocolumn,superscriptaddress,showpacs,longbibliography]{revtex4-1}

\usepackage{graphicx}
\usepackage{amssymb}
\usepackage{braket}
\usepackage{amsmath,mathrsfs}
\usepackage{color}

\newcommand{\black}{\color[rgb]{0.0,0.0,0.0}}

\begin{document}

\title{Optimal experimental demonstration of error-tolerant quantum witnesses}
\author{Kunkun Wang}
\affiliation{Department of Physics, Southeast University, Nanjing 211189, China}
\author{George C. Knee}
\email{gk@physics.org}
\affiliation{Department of Physics, University of Warwick, Gibbet Hill Road, Coventry CV4 7AL, United Kingdom}
\author{Xiang Zhan}
\affiliation{Department of Physics, Southeast University, Nanjing 211189, China}
\author{Zhihao Bian}
\affiliation{Department of Physics, Southeast University, Nanjing 211189, China}
\author{Jian Li}
\affiliation{Department of Physics, Southeast University, Nanjing 211189, China}
\author{Peng Xue}
\email{gnep.eux@gmail.com}
\affiliation{Department of Physics, Southeast University, Nanjing 211189, China}
\affiliation{State Key Laboratory of Precision Spectroscopy,
East China Normal University, Shanghai 200062, China}

\begin{abstract}
{Testing quantum theory on macroscopic scales is a longstanding challenge that might help to revolutionise physics. For example, laboratory tests (such as those anticipated in nanomechanical or biological systems) may look to rule out macroscopic realism: the idea that the properties of macroscopic objects exist objectively and can be non-invasively measured. Such investigations are likely to suffer from i) stringent experimental requirements, ii) marginal statistical significance and iii) logical loopholes. We address all of these problems by refining two tests of macroscopic realism, or `quantum witnesses', and implementing them in a microscopic test on a photonic qubit and qutrit.  The first witness heralds the invasiveness of a blind measurement; its maximum violation has been shown to grow with the dimensionality of the system under study. The second witness heralds the invasiveness of a generic quantum channel, and can achieve its maximum violation in any dimension -- it therefore allows for the highest quantum signal-to-noise ratio and most significant refutation of the classical point of view.  }
\end{abstract}
\maketitle
\section{Introduction}
Since the birth of quantum mechanics it has been difficult to reconcile the principle of quantum superposition with the intuitive experience of macroscopic objects, which appear to always inhabit explicit states independently of observation. Leggett and Garg~\cite{LeggettGarg1985,LeggettGarg1987,Leggett2008} formulated a possible solution by defining macroscopic realism (MR), a world view combining two assumptions: (MRps) Macroscopic realism \emph{per se} (that a macroscopic object will inhabit exactly one of its possible states at all times) and (NIM) Non-invasive measurability (that the object is not influenced by appropriately careful measurements). From these assumptions, they derived Leggett-Garg (LG) inequalities, which are used to test for the quantum behavior of a system undergoing coherent evolution~\cite{LeggettGarg1985,AharonovVaidman1990,HuelgaMarshallSantos1996,AvisHaydenWilde2010,Barbieri2009,DeviKarthikSudha2013,MaroneyTimpson2014,MoreiraKellerCoudreau2015,Emary2013}. The LG inequalities have been tested for a wide range of quantum mechanical systems, such as defect centers in diamond~\cite{WaldherrNeumannHuelga2011,GeorgeRobledoMaroney2013}, superconducting circuits~\cite{Palacios-LMalletNguyen2010,GroenRisteTornberg2013,KneeKakuyanagiYeh2016}, photons~\cite{XuLiZou2011,DresselBroadbentHowell2011,ZhouHuelgaLi2015,SuzukiIinumaHofmann2012,WangEmaryZhan2017}, atoms in optical lattices~\cite{RobensAltMeschede2015}, nuclear magnetic resonance~\cite{AthalyeRoyMahesh2011,KatiyarShuklaRao2013,KatiyarBrodutchLu2016} and phosphorus impurities in silicon~\cite{KneeSimmonsGauger2012}, and violations have been observed as quantum theory predicts. {  Alongside the major experimental challenge of applying the test to macroscopic systems (which could then in principle have a bearing on dynamical collapse theories~\cite{BassiLochanSatin2013}), proof-of-principle studies have so far concentrated on the implementation of noninvasive measurements. Approaches include using weak~\cite{Palacios-LMalletNguyen2010,GogginAlmeidaBarbieri2011} null-result~\cite{KneeKakuyanagiYeh2016,RobensAltEmary2016,WangEmaryZhan2017} or quantum non-demolition measurements~\cite{BudroniVitaglianoColangelo2015}; but latterly efforts have focussed on determining and accounting for measurement clumsiness with control experiments~\cite{Leggett1988,WildeMizel2011,GeorgeRobledoMaroney2013,KneeKakuyanagiYeh2016,KatiyarBrodutchLu2016}. } Furthermore, improvements to the protocol have been sought: for example, an alternative test of MR described, variously, as a quantum witness, no-signalling in time~\cite{KoflerBrukner2013,Halliwell2016,ClementeKofler2016}, or non-disturbance condition~\cite{KneeSimmonsGauger2012,GeorgeRobledoMaroney2013,MaroneyTimpson2014}. Compared to the original LG test which needs to involve at least three possible measurement times and the measurement of two-time correlations, the quantum-witness test enjoys many advantages: Because only instantaneous expectation values are required, it can usually be violated for a much wider parameter regime~\cite{ClementeKofler2016,KoflerBrukner2007,Bacciagaluppi2014}, and is more robust to imperfections~\cite{KneeKakuyanagiYeh2016}. Furthermore, it was recently shown that Fine's theorem (derived initially for local realism) does not apply for MR~\cite{ClementeKofler2016}. Notwithstanding the argument of Ref~\cite{Halliwell2016}, which offers a different perspective involving quasi-probabilities, LG inequalities do not form an optimal tight boundary for MR. 
In contrast, the quantum witness condition~\cite{LiLambertChen2012} is both necessary and sufficient for MR~\cite{ClementeKofler2016}. { There are however, additional improvements that may be made: as we show, a larger `quantum signal-to-noise ratio' may be had, and logical loopholes may be narrowed by altering the experimental protocol.} Here we continue in the pursuit of rigorous and amenable protocols for testing MR, reporting a microscopic experimental demonstation with a view to spurring-on tests which will non-trivially constrain future theories of physics.

{  Maroney and Timpson classified MRps into three types~\cite{MaroneyTimpson2014}. Two out of these three, termed Operational-Eigenstate-Support and Supra-Eigenstate-Support, involve hidden variables in a non-trivial way and cannot be ruled out by any experimental test on a two-level system~\cite{AllenMaroneyGogioso2016}. However, a third type, termed Eigenstate-Mixture macrorealism, which has it that all superpositions are in fact statistical mixtures in a preferred basis, can be ruled out by violating the LG inequality or a quantum witness condition. It is therefore this notion of MRps that we adopt for the remainder of this paper.}

\section{First witness}
Consider two observables, $A$ and $B$, measured at times $t=0$ and $t=T>0$, respectively. The measurement of observable $A$ is a blind measurement (the measurement is performed but the result is not recorded) shown in Fig.~1(a). The outcomes of the first measurement (of $A$) are written $\{a_i\}$ for $i=1,...,M$, with corresponding probabilities $P(a_i)$; $b$ is a particular outcome of the later measurement (of $B$). Based on the joint measurement of these two observables the probability of obtaining result $b$ in the later measurement is $P'(b)=\sum_{i=1}^M P(b|a_i)P(a_i)$ with $P(b|a_i)$ the conditional probability of the outcome $b$ given the earlier result $a_i$. The probability of outcome $b$ without the prior measurement of $A$ is written $P(b)$.
The first of our two quantum witnesses (derived in Refs. \cite{LiLambertChen2012,ClementeKofler2015} and elsewhere), is defined as~\cite{SchildEmary2015}
\begin{equation}
W:=P(b)-P'(b).
\label{eq:witness}
\end{equation}
Based on the tenets of MR, the presence of the blind measurement of $A$ should not affect the subsequent evolution of the system. One then has
\begin{equation}
W=0,
\label{eq:equality}
\end{equation}
our first quantum witness condition. {It can be derived under the same assumptions as the LG inequalites~\cite{LeggettGarg1987,KneeKakuyanagiYeh2016}. Equation (\ref{eq:equality}) can be violated by a quantum-mechanical system; the theoretical upper bound on the violation is given by~\cite{SchildEmary2015}
\begin{equation}
W_{\rm max}=1-\frac{1}{M},
\label{eq:max}
\end{equation}
where the number of blind-measurement outcomes $M\leq N$, the dimension of the system under study. The maximum violation can be found in the von Neumann measurement limit when $M=N$ and $W_{\rm max}=1-1/N$.

\begin{figure*}
   \includegraphics[width=.95\textwidth]{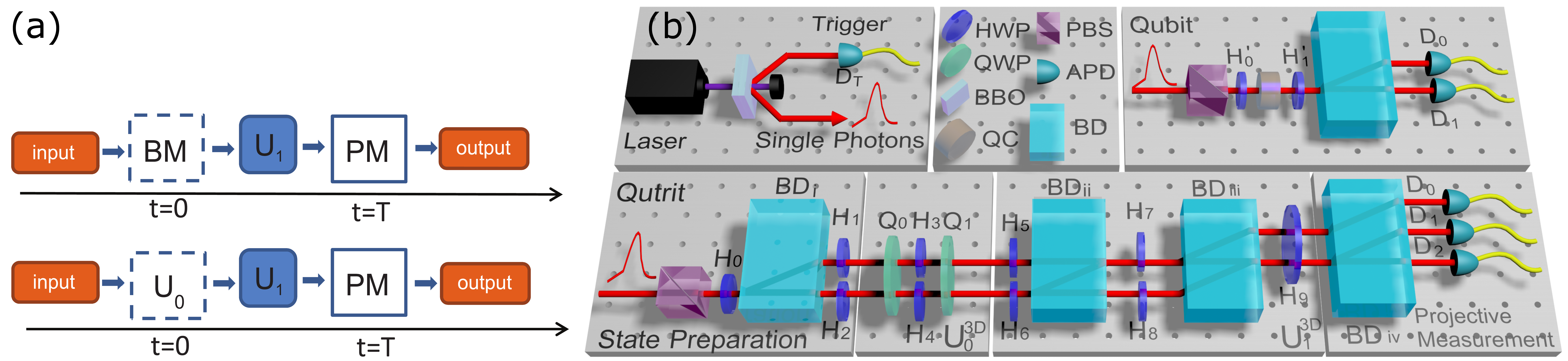}
   \caption{(a) Generic procedure for testing our two quantum-witnesses. We need four processes, including state preparation, state evolution (e.g. $\Phi(\rho)=U_1\rho U_1^\dagger$) and projective measurement (PM) of $B$. We also require the optional application of either i) blind measurement (BM) of $A$ (for testing witness $W$) or ii) some generic channel $\mathcal{E}(\rho)$, here chosen as a phase modulation $U_0\rho U_0^\dagger$ (for testing witness $V$). (b) Experimental setup. The heralded single photons are created via type-I SPDC in a BBO crystal and are injected into the optical network. Qubit states are prepared via a polarizing beam-splitter (PBS) and half-wave plates (HWPs). The phase channel $U^{2D}_0$ can be implemented by a set of wave plates (not shown), while a quartz crystal (QC) is used to realize a blind measurement of $A$. For the qutrit, the polarizing beam-splitter, half-wave plate and beam displacer are used for state preparation. $U^{3D}_0$ is implemented with half-wave plates and quarter-wave plates (QWPs), the blind measurement again with the quartz crystal (not shown), and $U^{3D}_1$ can be implemented with a cascaded interferometer network. The projective measurement of $B$ is realized via a beam displacer which maps the basis states of the qubit/qutrit to distinct spatial modes. 
}
\label{setup}
\end{figure*}

\section{Second witness}
{We propose a further improvement to LG's approach. Our second quantum witness $V$ consists in replacing the blind measurement (above) with a generic quantum channel:
\begin{equation}
V:=P(b)-P''(b).
\label{eq:witness2}
\end{equation}
$P''(b)$ denotes the probability of getting outcome $b$ in the later measurement of $B$ at $t=T$, when a generic quantum channel has been performed at $t=0$. Since the generic channel could be, but need not be, a blind measurement, we see that $P'(b)$ is a special case of $P''(b)$ when such an operation is chosen: see Fig.~\ref{setup}(a). 
 {To obtain a classical condition on $V$, consider the measurably zero (or near zero) effect of a general channel on certain `fixed point' preparation states.} Quantum theory predicts that a modulation of the global quantum phase of a state, for example, should have such a null effect. Attempting to interpret a superposition of such fixed-point states (where, according to quantum mechanics the various states now pick up definite relative phases) as a mere classical or incoherent mixture { (i.e. our assumption of MRps)} leads to the condition
\begin{equation}
V=0.
\label{eq:equality2}
\end{equation}

According to quantum theory, once the channel is relaxed from a blind measurement to some general map $\mathcal{E}$, we have {\black$V_\rho=\rm{Tr}(\Pi^b\left[\Phi(\rho)-\Phi(\mathcal{E}(\rho))\right])$ for $\Pi^b$} the projector corresponding to outcome $b$ of the later measurement, and $\Phi$ describing the time evolution of density operator $\rho$ from $t=0$ to $t=T$. It is easy to see that for pure state $\rho=|\psi\rangle\langle\psi|$, $\mathcal{E}(\rho)=U_0\rho U_0^\dagger$, and choosing $\Pi^b=\Phi(\rho)=U_1\rho U^\dagger_1$ ($U_0,U_1$ are unitary operators), we can achieve
\begin{equation}
V_{\rho}=1-|\langle \psi|U_0|\psi\rangle|^2.
\label{eq:witness2}
\end{equation}
Therefore, if $U_0$ orthogonalises $|\psi\rangle$, the $V$ witness can reach its algebraic maximum ($V_{\rm max}=1$) in any dimension. To achieve this maximum in general, (as can be seen by inspecting Eq.~(\ref{eq:witness2})) the channel must apply a pure phase to each fixed-point state, such that when a superposition of these states is injected the phases combine to form an orthogonal superposition. A blind measurement is equivalent to a random phase channel: this is at the root of the more modest maximum violations shown in~\eqref{eq:max}. 

{  The reader may object that the assumption of non-invasive measurability (NIM) has become non-invasive operability (NIO): namely, that a suitably careful operation (that need not be interpreted as a `measurement') can be made without affecting the future evolution of the system. 
By empirically testing for invasiveness, however, we shall see that in the end such auxiliary assumptions will play no role in the conditions that we test; so it is not necessary to consider the subtleties associated with replacing `measurement' with `operation'.}

\section{Error tolerance}The conditions $W=0$ and $V=0$ are not suitable for experimental test: since almost any real experiment will find a violation (because of finite statistical or technical errors), one may question the meaning of inferences made from data recorded in such an experiment.  It is better to construct the compound conditions
\begin{eqnarray}
&\min_i(W_i)\leq W_\sigma\leq \max_i(W_i)
\label{1stcondition}\\
&\min_i(V_i)\leq V_\sigma\leq \max_i(V_i),
\label{2ndcondition}
\end{eqnarray}
where the $W_{i}$ ($V_{i}$) correspond to the witness measured in the fixed-point states $|\psi_i\rangle$ { (which constitute a preferred basis)}, and $W_\sigma$ ($V_\sigma$) is a witness measured in the state $\ket{\psi_\sigma}=\sum_i\alpha_i\ket{\psi_i}$, where $\sum_i|\alpha_i|^2=1$. That the $W_i$ ($V_i$) can be nonzero constitutes an important quality of error-tolerance not present in conditions (\ref{eq:equality}) or (\ref{eq:equality2}): one requires the signature of quantumness to be significant w.r.t the control quantities $W_i$ or $V_i$. 
Recalling our interpretation of MRps as Eigenstate-Mixture macrorealism, the use of min and max functions to bound $W_\sigma$ and $V_\sigma$ allows for the most general MRps explanation interpreting the superposition $\sigma$ as an arbitrarily weighted incoherent mixture of the fixed-point states~\cite{KneeKakuyanagiYeh2016}. {  We therefore used only MRps (without using NIM or NIO) to derive (\ref{1stcondition}) and (\ref{2ndcondition}) and it is this assumption alone that is being tested.

Interestingly, once these conditions are adopted as necessary and sufficient conditions for MRps, the focus has shifted from the intrinsic properties of $W_\sigma$ or $V_\sigma$ to their relation to the $W_i$ or $V_i$ (respectively). Testing MRps is then merely asking the question, `does a certain measured quantity for preparation $\sigma$ lie in the convex hull of the same quantities measured for the classical states $\psi_i$?'. Furthermore, it is not even required that the $W_i (V_i)$ are small in magnitude, although this may be expedient for large violations.} In that case, it may be tempting to assume them to be zero when quantum theory predicts: but measuring the control quantities experimentally instead has the advantage of forming a more logically watertight argument \emph{contra} macrorealism, since the degree of `clumsiness'~\cite{WildeMizel2011} is determined and need not be assumed. Otherwise, the idea that the object under study is not quantum at all, but merely a classical system subjected to clumsy operations, remains a substantial loophole.

Assuming ancillary tests have been performed to determine $W_{i}$ ($V_{i}$), and found each to be (for example) close to zero, all that remains is the testing of $W_\sigma$ ($V_\sigma$). By judiciously choosing $U_0$ and $\ket{\psi_\sigma}$, the maximum violation of (\ref{2ndcondition}) is generally greater than the maximum violation of (\ref{1stcondition}), and furthermore independent of the dimension of the system. This is clearly more experimentally favourable; there is a greater robustness to imperfection, and violations will emerge sooner from statistical noise, i.e. with fewer experimental trials.} Note that with the introduction of the control quantities $V_i$, it might seem that a violation of (\ref{1stcondition}) or (\ref{2ndcondition}) with magnitude greater than one is possible -- e.g. by arranging all $V_i=-1$ and $V_\sigma=+1$. Such violations are, however, not permitted by any deterministic operation allowed in quantum mechanics (as we show in Appendix \ref{maxviolation}). 
{Next we will give details and results of an experimental demonstration of violation of these witness conditions.}
}

\section{Optimal violations} It is simple to find preparations, channels and measurements which saturate the maximum violations of our two witnesses. In our experimental demonstration, we choose the fixed-point states to be an orthonormal basis $\{|\psi_0\rangle,|\psi_1\rangle\}$ for the qubit and $\{|\phi_0\rangle,|\phi_1\rangle,|\phi_2\rangle\}$ for the qutrit, which all satisfy $W_i=V_i=0$ because they form the eigenvectors of i) the blind measurement observable $A$, and ii) of the unitary channel $U_0$. These basis states will be encoded in a combination of polarization and path degrees of freedom of single photons (see Fig.~\ref{setup}(b)). Writing matrices in these bases, we choose 
\begin{equation}
U^{2D}_0=\left(
\begin{array}{cc}
$1$&$0$\\$0$&$-1$
\end{array}
\right)
\quad\rm{and}\quad
U^{2D}_1=\left(
\begin{array}{cc}
1&1\\1&-1
\end{array}
\right)
\end{equation}
  for the qubit, and
\begin{align}
U^{3D}_0&=\left(
\begin{array}{ccc}
1&0&0\\0&\rm{e}^{i\frac{2\pi}{3}}&0\\0&0&\rm{e}^{i\frac{4\pi}{3}}
\\\end{array}\right)\quad\rm{and}\quad\\
U^{3D}_1&=\left(
\begin{array}{ccc}
                                    \sqrt{\frac{2}{3}} & \frac{\sqrt{6}}{6} & -\frac{\sqrt{6}}{6} \\
                                    0 & \frac{\sqrt{2}}{2} & \frac{\sqrt{2}}{2} \\
                                    \sqrt{\frac{1}{3}} & -\sqrt{\frac{1}{3}} & \sqrt{\frac{1}{3}} \\
                                 \end{array}\right)
\end{align}
for the qutrit. The measurement operator is chosen as { $\Pi^b$}$=|\psi_1\rangle\langle \psi_1|$ or $|\phi_2\rangle\langle \phi_2|$ respectively. The superposition states are $\ket{\psi_\sigma}=(\ket{\psi_0}-\ket{\psi_1})/\sqrt{2}$ and $\ket{\phi_\sigma}=(\ket{\phi_0}-\ket{\phi_1}+\ket{\phi_2})/\sqrt{3}$, which achieve $W_\sigma^{2D}=1/2,W_\sigma^{3D}=2/3,V_\sigma^{2D}=V_\sigma^{3D}=1$.

\begin{figure*}
   \includegraphics[width=.95\textwidth]{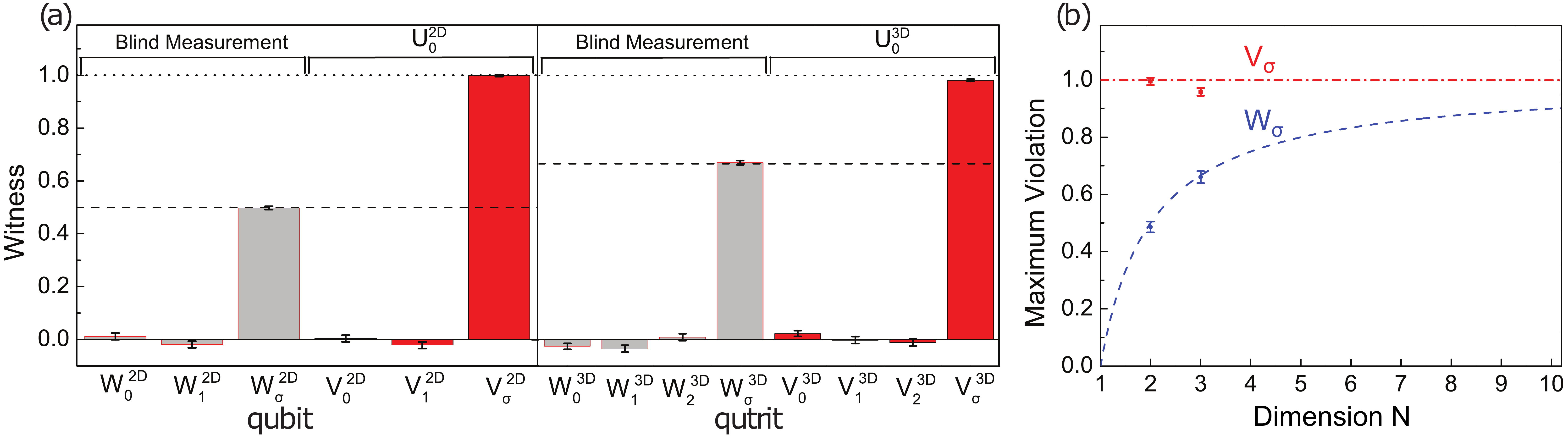}
   \caption{(a) Experimentally determined values for our two quantum witnesses $W$ and $V$, for the fixed-point preparations $0$ and $1$ (also $2$) and for the superposition preparation $\sigma$ of a photonic qubit (and qutrit). Theoretical predictions ate represented by dashed and dotted lines. Error bars indicate the statistical uncertainty which is obtained based on assuming Poissonian statistics. (b) Experimental results showing maximum violations of both quantum witness conditions for two-level and three-level photonic systems. The blue (red) dashed (dashed-dotted) line represents the theoretical predictions of the maximum violations of the first (second) quantum witness condition.
   }
\label{two}
\end{figure*}



\section{Experimental results} The state of the qubit can be represented by two orthogonal polarization states of single photons, which are generated via a type-I spontaneous parametric down-conversion (SPDC)~\cite{KwiatWaksWhite1999} -- see Fig. 1(b). The polarization-degenerate photon pairs at a wavelength of $801.6$nm are produced using a $0.5$mm-thick $\beta$-barium-borate (BBO) crystal pumped by a $90$mW diode laser, which is filtered out by a $3$nm bandpass filter. The signal photon is heralded for evolution and measurement by detection of a trigger photon: coincidences are registered by avalanche photodiodes with a 7ns time window. A polarizing beam splitter (PBS) and a half-wave plate (H'$_0$), set at certain angles, are used to prepare the input states. $U^{2D}_1$ is realized by a half-wave plate (H'$_1$) at $22.5^\circ$. A birefringent calcite beam displacer is used to map the basis states of the qubit to two spatial modes and to accomplish the projective measurement of $B$. The probability of the photons being measured in $\ket{\psi_1}$ is estimated by normalizing photon counts in each spatial mode to total photon counts. The count rates are corrected for differences in detector efficiencies and losses before the detectors. We assume that the ensemble of prepared photons is fairly represented by the sample of detected photons (fair sampling assumption).
Total coincidence counts are about $13,000$ over a collection time of $6$s. 

For a qutrit, the basis states $\ket{\phi_0}$, $\ket{\phi_1}$, and $\ket{\phi_2}$ are encoded by the horizontal polarization of the photon in the upper
mode, and the horizontal/vertical polarization of the photon in the lower mode. After passing through a polarizing beam-splitter and half-wave plate (H$_{0}$), the heralded single photons pass a beam displacer (BD$_1$) followed by two half-wave plates (H$_{1}$ and H$_{2}$)~\footnote{The optical axis of the beam displacer is cut so that vertically polarized light is directly transmitted and horizontal light undergoes a $3$mm lateral displacement into a neighboring spatial mode.}, one in each spatial mode \cite{ZhanZhangLi2016}. By tuning their angles, the photons are prepared in one of the input states. The unitary channel $U^{3D}_1$ (belonging to $SU(3)$) can be decomposed into three unitary channels, each of which applies a rotation to just two of the basis states, leaving the other unchanged. Each of them can be realized by two half-wave plates and a beam displacer~\cite{WangZhanBian2016}. One of the half-wave plates is used to apply a rotation on two modes of the qutrit state and the other is used to compensate the optical delay. The beam displacers are used to translate between polarization encoding and spatial-mode encoding. In this way, wave plates acting on the two polarization modes propagating in the same spatial mode can be used to effect two-(spatial)-mode transformations. See Appendix \ref{moredetails} for further details. {  In Appendix \ref{arbitraryN}, we present a scheme that generalises this approach to any desired dimension $N$. }

To implement the blind measurement, note that after it is applied, the system is a mixture with diagonal density operator $\rho_{\rm{blind}}=\Gamma(\rho) = \sum_m \langle \psi_m|\rho|\psi_m\rangle \ket{\psi_m}\bra{\psi_m}$. Note that $\Gamma(\rho)$ is a completely dephasing channel. Thus the blind measurement can be realized by a quartz crystal, inserted into the lower spatial mode so as to destroy the spatial coherence of the photons. The coherence length of the photons is $L_c\approx\lambda^2/\Delta\lambda$, where $\lambda$ is the central wavelength of the source and $\Delta\lambda$ is the spectral width of the source \cite{AkcayParreinRolland2002}. Hence the thickness of the quartz crystal should be at least $23.97$mm: In our experiment, it is about $28.77$mm. To test the violation of our second quantum witness condition (\ref{2ndcondition}), we replace the quartz crystal by wave plates with certain setting angles which are used to realize the unitary evolution $U^{2D}_0$ or $U^{3D}_0$.

In Fig.~2(a), the experimentally determined values of our witnesses are shown. Due to the high precision nature of our laboratory setup, we found all quantities to be close to their predicted values. In particular the fixed point preparations gave witness values close to zero, and we found $W^{2D}_\sigma=0.4980\pm0.0060\, (35\rm{sd}), V^{2D}_\sigma=0.9998\pm0.0004 \,(80\rm{sd}),W^{3D}_\sigma=0.6700\pm0.0080\,(44\rm{sd}), V^{3D}_\sigma=0.9820\pm0.0020\,(72\rm{sd})$. The number of standard deviations (sd) of violation, given by $(\Xi_\sigma-\max_i\Xi_i)/\sqrt{\rm{Var}(\Xi_\sigma)+\rm{Var}(\max_i\Xi_i)}$ ($\Xi=W, V$), is shown parenthetically. Note how using the second witness lead to violations of MRps with higher statistical significance. In Fig.~2(b), these data are shown alongside theoretical predictions for the maximum violations of both witnesses. 

\section{Conclusion}{ Though violation of the LG inequality has become the standard laboratory proof of `quantumness', ruling out MR in the same way that violation of Bell's inequality rules out local realism, the LG inequality is only necessary (but not sufficient) for MR. In this paper, we employed the necessary and sufficient quantum-witness condition, which allows for greater statistical significance given fewer experimental resources. We significantly increase the statistical significance yet further, by changing from a blind measurement to a unitary channel. In both cases, we measured the classical disturbance introduced by our operations, {  which tightens the logical loopholes in the demonstration, as well as removing the need to assume any kind of non-invasiveness}. We have recorded experimental violations in both a photonic qubit and qutrit. Our results agree well with the theoretical maximum violations, and showcase perhaps the final, fine tuned, protocol for testing macroscopic realism {  \emph{per se}}, which gives the greatest possible chance for finding convincing violations in truly macroscopic systems, while remaining error-tolerant and evading the clumsiness loophole.

\section*{Acknowledgments}
We would like to thank Neill Lambert and Clive Emary for helpful discussions. We acknowledge support by NSFC (Nos.~11474049 and 11674056) and NSFJS (No. BK20160024). G. C. K. is supported by the Royal Commission for the Exhibition of 1851.

\appendix
\section{Maximum violation of the witnesses}
\label{maxviolation}
Our second witness is defined
\begin{eqnarray}
V_\rho:=\rm{Tr}(\Pi^{b\prime}(\rho_\sigma-\mathcal{E}(\rho_\sigma))),
\end{eqnarray}
where $\Pi^{b\prime} =\Phi^\dagger(\Pi^b)$ 
 is related to a measurement in the preferred basis $\Pi^{b\prime}$ by the channel $\Phi$. Clearly
\begin{eqnarray}
-1\leq V_\rho\leq 1
\end{eqnarray}
because it is the difference of two probabilities. Now, by using the completeness relation $\sum_i |\psi_i\rangle\langle \psi_i |=\mathbb{I}$ for an orthonormal set of basis vectors, one can see that
\begin{eqnarray}
\sum_i V_i &= \rm{Tr}\left(\Pi^{b\prime}\left[ \sum_i |\psi_i\rangle\langle \psi_i |- \mathcal{E} \left(\sum_i |\psi_i\rangle\langle \psi_i |\right) \right]\right)\nonumber\\
&= \rm{Tr}\left(\mathbb{I}\left[\Pi^{b\prime}-\mathcal{E}^\dagger(\Pi^{b\prime})\right]\right)\nonumber\\
&=0.
\end{eqnarray}
Above we used the trace-preserving property of $\mathcal{E}$. Therefore we have that $\max_i V_i\geq 0$ and $\min_iV_i\leq 0$, and thus
\begin{eqnarray}
V_\sigma - \max_i V_i\leq+1&\\
V_\sigma - \min_i V_i\geq -1&,
\end{eqnarray}
and the maximum violation of our condition has a magnitude of unity. Since $\mathcal{E}$ is any quantum channel, it could be the completely dephasing channel. So this proof also establishes
\begin{eqnarray}
W_\sigma - \max_i W_i\leq+1&\\
W_\sigma - \min_i W_i\geq -1&,
\end{eqnarray}
although tighter bounds were proved by Schild and Emary~\cite{SchildEmary2015}. We have made the assumption that $\mathcal{E}$ is a quantum channel; that is, it belongs to the subset of quantum operations that preserve the trace of the density operator. Trace-decreasing operations are indeterministic and are only possible via post-selection. We leave the discussion of maximum violations for that complementary subset, and their interpretation, for future work.


\section{Experimental realization of the unitary evolutions and projective measurements for two- and three-level systems}
\label{moredetails}
To test the violation of first quantum witness condition, we insert a quartz crystal to destroy spatial coherence. Whereas, to test the violation of second quantum witness condition, we replace the quartz crystal with a set of wave plates which are used to realize the unitary channels $U^{2D}_0$ or $U^{3D}_0$. For implementation of $U^{2D}_0$, we use a half-wave plate to realize the Pauli operator $\sigma_z$ on the qubit state. For a qutrit case, the evolution $U^{3D}_0$ is a channel that adds a phase $e^{i4\pi/3}$ on the state $\ket{\phi_2}$, $e^{i2\pi/3}$ on $\ket{\phi_1}$ and keeps $\ket{\phi_0}$ unchanged, and can be realized by wave plates with certain setting angles inserting to the proper spatial modes (upper and lower modes).

In the projective measurement stage, a beam displacer is used to map the basis states of qutrit to three spatial modes and to accomplish the projective measurement $\Pi^b=\ket{\phi_2}\bra{\phi_2}$. For photon detection, we record pair-wise coincidences between the heralding detector and any one of the three detectors in the measurement apparatus (heralded single clicks). Simultaneous registrations in the heralding detector and two of the three detectors for measurement give the three-body coincidences. Such coincidences might occur when the source generates more than one photon per pulse, but we find the frequency of these to be negligible in our experiment.


We estimate the relative detection efficiency of each detector $D_i$ assuming that the sum of the photon counts at the detectors does not depend on the configuration of optical components before the detector. We then use these efficiencies to correct each count rate $\tilde{N_i}=N_i/D_i$. For the probability of detection we then have $P_i=\tilde{N_i}/\sum^3_{i=0,1,2}\tilde{N_i}$, where $\tilde{N_i}$ is the corrected number of heralded clicks at detector $i$. The corrections are made to eliminate the effects of different efficiencies of the detectors.

\section{Generalization to arbitrary $N$}
\label{arbitraryN}
Just as shown in Fig. 1 (a) of the main text, there are four processes in this experiment, including state preparation, operation of $U_0$ (or blind measurement), state
evolution $U_1$ and projective measurement of $B$. In order to extend the dimension of the system to $N$-dimension, the primary problem is to find a setup to implement arbitrary unitary channels, since these can be combined to enable full state preparation and measurement (as well as $U_0$ and $U_1$). For the blind measurement, it can be realized by inserting quartz crystals with different thicknesses in different optical modes.

It has been proven that any $N \times N$ unitary array can be decomposed as a sequence of $U(2)$ transformations on two-dimensional subspaces of the $N$-dimensional Hilbert space, with the complementary subspace unchanged~{\cite{ReckZeilingerBernstein1994}}:
\begin{equation}
  U(N)=(E_{N,N-1} \cdot E_{N,N-2}...E_{2,1}\cdot S )^{-1}.
  \label{RZB}
\end{equation}
Here $E_{i,j}$ is an $N$-dimensional identity array with the elements of $E_{i,j}^{i,i}, E_{i,j}^{i,j}, E_{i,j}^{j,i}$, and $E_{i,j}^{j,j}$ replaced by the corresponding $U(2)$ array elements. $S$ is used to adjust the phase of the output.

Figure \ref{setupN} shows the experiment setup to achieve $N$-dimensional unitary evolution.  If $N$ is an even number, the basis states are encoded by the horizontal / vertical polarization in $n=N/2$ spatial modes, i.e.
$$
|1\rangle \& |2\rangle, |3\rangle \& |4\rangle,...,|N-1\rangle\&|N\rangle\sim \rm{H}_1\&\rm{V}_1,\rm{H}_2\&\rm{V}_2,...,\rm{H}_n\&\rm{V}_n.
$$ 
 Else if $N$ is odd, basis state $|1\rangle$  is encoded by the horizontal polarization in the first spatial mode, and the remaining basis states are encoded by the horizontal / vertical polarization in $(N-1)/2$ spatial modes: i.e.
 $$
|1\rangle,|2\rangle\&|3\rangle,...,|N-1\rangle\&|N\rangle\sim \rm{H}_1,\rm{H}_2\&\rm{V}_2,...,\rm{H}_n\&\rm{V}_n
$$
with $n = \frac{N+1}{2}$. According to Eq. (\ref{RZB}), if we can combine the different spatial modes into one spatial mode and apply arbitrary $U(2)$ in this mode in sequence, we can implement an arbitrary unitary operator across many spatial modes. An arbitrary transformation of the polarization of a photon can be realized by a set of wave plates which contain two half-wave plates (blue rectangle in Fig. \ref{setupN}) and one quarter-wave plate (orange rectangle). As shown in Fig. \ref{setupN}, the first $E_{N,N-1}$ can be achieved by inserting a set of wave plates in the last optical mode.  Then by applying a half-wave plate at $45^\circ$ in all of the other modes, after passing through the first beam displacer (BD$_1$), we can combine the levels of $N$ and $N-2$ together. By inserting another wave plate set, we can implement $E_{N,N-2}$. We can use the same procedure to realize the following $E_{i,j}$. After this setup, the output labelling is reversed. So that we have e.g. (H$_n'$\&V$'_n$),\dots,(H$'_2$\&V$'_2$),(H$'_1$\&V$'_1$) corresponding to the basis states of $(|N-1\rangle\&|N\rangle) \ldots (|3\rangle\&|4\rangle),(|1\rangle\& |2\rangle)$ . The maximum number of beam displacers needed to build this $N$-dimensional unitary operator is $2N-4$ when $N$ is an even number and $2N-3$ for odd. This number grows only polynomially with the number of dimensions.

In order to extract the maximum violation of our witnesses, it suffices to prepare the maximally coherent state~\cite{BaumgratzCramerPlenio2014} $|\psi_\sigma\rangle=\sum_i |\psi_i\rangle/\sqrt{N}$ and make the final measurement include $\Pi_b$ such that $U_1^\dagger \Pi_b U_1=|\psi_\sigma\rangle\langle \psi_\sigma|$ is a projector onto this state~\cite{SchildEmary2015}. For the $W$ witness, the blind measurement of course projects onto the $|\psi_i\rangle$ states; for the $V$ witness, one may use $U_0=\sum_{k=0}^N e^{i(k)2\pi/N}|\psi_k\rangle\langle \psi_k|$. In this way, a phase from each basis state contributes to the violation. It may be possible to design other schemes where only phases from e.g. macroscopically distinct basis states contribute: see Ref~\cite{LambertDebnathKockum2016} for a similar analysis of measurement schemes for the LG inequality.

\begin{figure*}[ht]
  \centering
  \includegraphics[width=\textwidth]{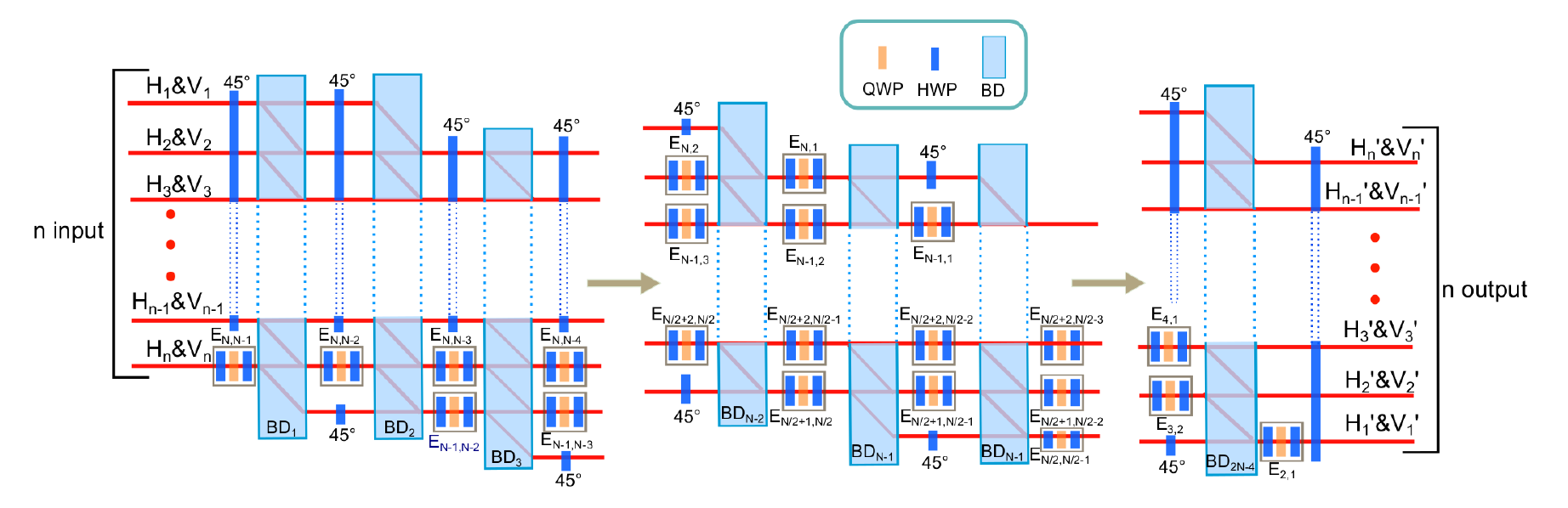}
  \caption{Experimental setup with cascaded interferometers which is able to implement any unitary operator. In this diagram, $N$ is assumed to be an even number.}
  \label{setupN}
\end{figure*}

%
%
\bibliography{witnessbib}

\end{document}